\documentclass[trackchanges]{aastex701}
\usepackage{amsmath, amssymb, amsfonts}
\usepackage{multirow}
\usepackage{tabularx}
\usepackage{booktabs} 
\usepackage{bm}

\begin{document}

\title{Influence of Finite-Nuclei Constraints on High-Density Transitions and Neutron Star Properties}

\author[orcid=0000-0002-0812-2702,sname='Venneti']{Anagh Venneti}
\affiliation{Department of Physics, Birla Institute of Technology and Science, Pilani, Hyderabad Campus, Jawahar Nagar, Kapra Mandal, Medchal District, Telangana 500078, India}
\email[show]{p20210060@hyderabad.bits-pilani.ac.in}  

\author[orcid=0000-0003-0221-3651, sname='Banik']{Sarmistha Banik} 
\affiliation{Department of Physics, Birla Institute of Technology and Science, Pilani, Hyderabad Campus, Jawahar Nagar, Kapra Mandal, Medchal District, Telangana 500078, India}
\email{sarmistha.banik@hyderabad.bits-pilani.ac.in}

\author[orcid=0000-0001-5032-9435, sname=Agrawal]{Bijay K Agrawal}
\affiliation{Saha Institute of Nuclear Physics, 1/AF Bidhannagar, Kolkata, 700064, West Bengal, India}
\affiliation{Homi Bhabha National Institute, Anushakti Nagar, Mumbai, 400094, Maharashtra, India}
\email{sinp.bijay@gmail.com}

\begin{abstract}
We construct posterior distributions of the equation of state (EoS) for matter beyond the inner crust of neutron stars by incorporating finite-nuclei (FN) constraints within relativistic mean-field models. These constraints are implemented in three complementary ways: (i) through theoretical bounds on the EoS, (ii) implicitly via nuclear matter parameters, and (iii) explicitly by enforcing consistency with experimental binding energies and charge radii of selected nuclei. The resulting low-density nucleonic EoSs are subsequently matched to a model-agnostic speed-of-sound parametrization, constrained by astrophysical observations, including NICER mass–radius measurements, tidal deformability limits from GW170817, and lower bounds on the maximum neutron-star mass inferred from radio pulsar observations. We find that the admissible range of the transition density is strongly sensitive to the choice of the low-density EoS. In particular, the inclusion of explicit FN constraints significantly reduces the allowed parameter space of the nucleonic EoS at low densities, narrowing the transition-density range by nearly a factor of two. Consequently, neutron-star properties inferred from EoSs with explicit FN constraints differ substantially, with especially pronounced effects for low-mass neutron stars and their correlations with nuclear matter parameters. A quantitative comparison, using metrics based on Mahalanobis distance, shows consistency of the explicit constraints with PSRs J0740+6620, J0030+0451, and J0437–4715, but suggest a possible tension with PSR J0614–3329. These findings underscore the critical importance of a consistent treatment of finite-nuclei properties for reliably inferring the behavior of high-density matter and the presence of possible phase transitions from astrophysical observations.

\end{abstract}



\section{Introduction} 

Neutron stars (NSs) provide a unique window into the properties of ultra-dense and highly isospin-asymmetric matter \citep{LATTIMER2007109, RevModPhys.89.015007, Baym_2018}. They reach densities far exceeding those achievable in present-day terrestrial experiments, making the study of their internal structure and macroscopic properties crucial for understanding nuclear matter under extreme conditions. At such densities, nucleon-rich hadronic matter is theoretically predicted to undergo transitions to exotic degrees of freedom, including hyperons \citep{glendenning2012compact, Banik_2014, PhysRevC.53.1416}, meson condensates \citep{roy2025signatureskcondensationneutron, Malik_2021}, or deconfined quark matter \citep{WEBER2005193, RevModPhys.80.1455}. The emergence of these components manifests as nontrivial features in the behavior of the equation of state (EoS).

Multiple observational channels now provide complementary constraints on NS properties. These include simultaneous mass--radius measurements from NICER observations \citep{miller2019psr, miller2021radius, riley2019nicer, riley2021nicer, Choudhury_2024, Mauviard_2025}, measurements of the dimensionless tidal deformability from the binary neutron star merger GW170817 \citep{abbott2017gw170817grav}, and lower bounds on the maximum mass inferred from radio timing of massive pulsars \citep{fonseca2021refined, doi:10.1126/science.1233232, Arzoumanian_2018}. These observables probe different density regimes of the EoS. For example, the radii of low-mass neutron stars are primarily sensitive to the low-density behavior, while tidal deformability and the maximum mass depend more strongly on the intermediate- and high-density regions.

The lack of direct experimental access to high-density matter necessitates the use of theoretical modeling to interpret these observations \citep{refId0, shirke2025psrj06143329nicercase, 10.3389/fphy.2024.1531475, yang2025strangestarsadmixedmirror, PhysRevD.110.063040}. A wide variety of theoretical models exist, differing in their assumptions regarding particle content, interactions, and many-body treatments. Numerous studies \citep{saha2025rolepsrj06143329defining, Li_2025, Huang_2025, bmsk-8n85, shahrbaf2025observationalprobesneutronstar, GUHAROY2024139128} have attempted to use NS observations to infer the transition density between hadronic and non-hadronic phases, employing first-order phase transitions, smooth crossovers, and other phenomenological constructions. However, these investigations often arrive at conflicting conclusions regarding the presence and nature of such transitions, reflecting their strong dependence on modeling assumptions and the choice of constraints imposed across different density regimes.

In our previous work \citep{venneti2024unraveling}, we demonstrated that the manner in which low-density finite-nuclei (FN) constraints are incorporated, implicitly or explicitly, has a substantial impact on predicted NS properties and their correlations with the model parameters. Explicitly constraining FN observables, such as binding energies and charge radii, which are known with relatively high precision, was shown to significantly restrict the low-density EoS. Despite this, the implications of explicit FN constraints for the transition density and the possible emergence of non-nucleonic degrees of freedom at higher densities have not been systematically explored. Given that low-density EoS features influence the permissible range of transition densities, these constraints are expected to play an important role in shaping the high-density behavior of neutron star matter.

In this work, we address this gap by employing an outer core EoS constructed within the relativistic mean-field (RMF) formalism and subjecting it to three different sets of constraints, which we label as theoretical, implicit, and explicit, within a Bayesian framework. Uncertainties at high densities, in the inner core, are accounted for using a minimal and relatively model-agnostic speed-of-sound parameterization ~\citep{Tews2018} beyond a transition density, \( \rho_{tr} \), which is treated as a free parameter. We investigate how the different sets of constraints influence the inferred transition density, the properties of the EoS, neutron star observables, and the resulting correlation structure. 

Before proceeding, we clarify the usage of the term ``transition density'' in this work. Throughout this manuscript, the term refers to a change in the effective behavior of the EoS beyond this density, rather than to a thermodynamic phase transition with a specified microscopic origin. Given the substantial uncertainties in the microscopic description of matter at high densities, we treat the transition density as a phenomenological indicator of the regime where a nucleonic model may no longer be adequate.

The paper is organized as follows: Section~\ref{RMF} describes the RMF formalism; the high-density speed-of-sound parameterization is presented in Section~\ref{HD}; the Bayesian framework and imposed constraints are detailed in Section~\ref{Bayesian}. Our results and discussion are given in Section~\ref{Results_And_Discussions}, followed by a summary and conclusions drawn in Section~\ref{Conclu}.

\section{Equation of state and the constraints} 
\subsection{Nucleonic Matter}
\label{RMF}
The equation of state (EoS) of charge-neutral, beta-equilibrated nucleonic matter is constructed within the framework of relativistic mean-field (RMF) theory. In this approach, nuclear matter consists of interacting nucleons whose dynamics are mediated by meson exchanges: the scalar $\bm{\sigma}$ meson provides short-range attraction, the vector $\bm{\omega}$ meson generates short-range repulsion, and the isovector $\bm{\rho}$ meson accounts for isospin-dependent interactions. In addition to mediating nucleon--nucleon interactions, the mesonic fields also exhibit nonlinear self-interactions (for the $\bm{\sigma}$ and $\bm{\omega}$ mesons) and cross-interactions between the $\bm{\omega}$ and $\bm{\rho}$ mesons. The effective RMF Lagrangian density~\citep{Dutra2014,Zhu_2023} is written as
\begin{equation}
\mathcal{L}_{\rm RMF}
=
\mathcal{L}_{\rm N}
+
\mathcal{L}_{\sigma}
+
\mathcal{L}_{\omega}
+
\mathcal{L}_{\rho}
+
\mathcal{L}_{\rm int},
\end{equation}
where the individual terms describe the nucleonic sector, mesonic kinetic and mass contributions, and interaction terms. The nucleon Lagrangian is given by
\begin{align}
\mathcal{L}_{\rm N} &=
\bar{\psi}\left(i\gamma^\mu \partial_\mu - m\right)\psi
+ g_\sigma \sigma \bar{\psi}\psi
- g_\omega \bar{\psi}\gamma^\mu \omega_\mu \psi
- \frac{g_\rho}{2} \bar{\psi}\gamma^\mu \vec{\rho}_\mu \cdot \vec{\tau}\psi ,
\end{align}
while the mesonic contributions take the form
\begin{align}
\mathcal{L}_{\sigma} &=
\frac{1}{2}\left(\partial^\mu \sigma \partial_\mu \sigma - m_\sigma^2 \sigma^2\right)
- \frac{A}{3}\sigma^3
- \frac{B}{4}\sigma^4 , \\
\mathcal{L}_{\omega} &=
-\frac{1}{4}\Omega^{\mu\nu}\Omega_{\mu\nu}
+ \frac{1}{2} m_\omega^2 \omega^\mu \omega_\mu
+ \frac{C}{4}\left(g_\omega^2 \omega_\mu \omega^\mu\right)^2 , \\
\mathcal{L}_{\rho} &=
-\frac{1}{4}\vec{B}^{\mu\nu}\!\cdot\!\vec{B}_{\mu\nu}
+ \frac{1}{2} m_\rho^2 \vec{\rho}_\mu \!\cdot\! \vec{\rho}^{\,\mu} .
\end{align}
The cross-interaction between the $\bm{\omega}$ and $\bm{\rho}$ mesons is introduced through
\begin{equation}
\mathcal{L}_{\rm int} =
\frac{1}{2}\Lambda_v g_\omega^2 g_\rho^2
\left(\omega_\mu \omega^\mu\right)
\left(\vec{\rho}_\mu \cdot \vec{\rho}^{\,\mu}\right).
\end{equation}
Here,
$\Omega_{\mu\nu} = \partial_\mu \omega_\nu - \partial_\nu \omega_\mu$
and
$\vec{B}_{\mu\nu} = \partial_\mu \vec{\rho}_\nu - \partial_\nu \vec{\rho}_\mu
- g_\rho (\vec{\rho}_\mu \times \vec{\rho}_\nu)$
denote the field tensors of the $\bm{\omega}$ and $\bm{\rho}$ mesons, respectively.
The nucleon mass is denoted by $m$, while $m_\sigma$, $m_\omega$, and $m_\rho$
represent the meson masses.

The coupling constants of the RMF model are determined by nuclear matter
parameters (NMPs) evaluated at the saturation density $\rho_0$, including the
binding energy per nucleon $E_0$, incompressibility $K_0$, skewness parameter
$Q_0$, effective mass ratio $m^*/m$, symmetry energy $J_0$, and its slope $L_0$. $\rho_0$, $E_0$, $K_0$, $Q_0$ and $m^*/m$ are labeled as isoscalar NMPs, while $J_0$ and $L_0$ are the isovector NMPs. These nuclear matter parameters uniquely fix the couplings
$g_\sigma$, $g_\omega$, $g_\rho$, $A$, $B$, $C$, and $\Lambda_v$
\citep{Dutra2014,Zhu_2023}. The EoS is constructed within this framework for densities above $0.5\,\rho_{\text{ref}}$ till $\rho_{tr}$, where $\rho_{\text{ref}} = 0.16\, \text{fm}^{-3}$. 

As emphasized earlier, $\rho_{tr}$ denotes the density beyond which additional EoS flexibility is introduce, rather than a thermodynamic phase transition. The crustal region is  the Baym--Pethick--Sutherland (BPS) EoS for the outer crust~\citep{baym1971ground} and the Negele--Vautherin (NV) EoS for the inner crust~\citep{Negele1973}.
\subsection{High-Density Matter}
\label{HD}

At supranuclear densities, such as those realized in neutron star interiors,
additional degrees of freedom or exotic phases of matter may emerge.
Possible scenarios include deconfined quark matter~\citep{WEBER2005193},
hyperonic matter~\citep{PhysRevC.53.1416, Banik_2014},
kaon condensation~\citep{roy2025signatureskcondensationneutron},
and other non-nucleonic phases~\citep{thakur2024towards}.
To capture these possibilities in a model-independent manner,
the EoS is extended beyond a transition density $\rho_{\rm tr}$
using a parameterized speed-of-sound prescription~\citep{Tews2018},
allowing for smooth crossover-like behavior. The squared speed of sound $\frac{c_s^2}{c^2}$, as a function of number density $\rho$, is parameterized as
\begin{align}
\frac{c_s^2}{c^2} &=
\frac{1}{3}
- c_1 \exp\!\left[-\frac{(\rho - c_2)^2}{n_{\rm bl}^2}\right]
+ h_p \exp\!\left[-\frac{(\rho - n_p)^2}{w_p^2}\right]
\left[1 + \operatorname{erf}\!\left(s_p \frac{\rho - n_p}{w_p}\right)\right] .
\end{align}
Here, $c$ indicates the speed of light. The parameters $c_1$ and $c_2$ are fixed by enforcing continuity of the
speed of sound and its first derivative at $\rho_{\rm tr}$.
The remaining parameters (labeled HDPs) $n_{\rm bl}$, $h_p$, $n_p$, $w_p$, and $s_p$
control the width, amplitude, location, and asymmetry of the sound-speed peak respectively and are sampled from uniform prior distributions following Ref.~\citep{Tews2018}.

\subsection{Bayesian Framework}
\label{Bayesian}

Constraints on the EoS are imposed using Bayesian inference.
Bayes' theorem relates the posterior distribution of the model parameters
$\bm{\Theta}$ to the observational data $D$ as
\begin{equation}
P(\bm{\Theta} \mid D) =
\frac{\mathcal{L}(D \mid \bm{\Theta}) P(\bm{\Theta})}{\mathcal{Z}},
\end{equation}
where $P(\bm{\Theta})$ denotes the prior,
$\mathcal{L}(D \mid \bm{\Theta})$ is the likelihood,
and $\mathcal{Z}$ is the Bayesian evidence. The parameter set $\bm{\Theta}$ includes the NMPs
$(\rho_0, E_0, K_0, Q_0, m^*/m, J_0, L_0)$
along with the HDPs
$(n_{\rm bl}, h_p, w_p, n_p, \rho_{\rm tr})$
that govern the speed-of-sound profile. The priors considered for the parameters are shown in Table~\ref{tab:priors_only}.
\begin{table}[h!]
\centering
\caption{\label{tab:priors_only}
Priors adopted for the nuclear matter parameters (NMPs), high-density parameters (HDPs), and the transition density.
Uniform priors are denoted by U(low, high). $\rho_{\text{ref}}$ is the reference density of $0.16~\mathrm{fm}^{-3}$.
}
\begin{tabular}{c c c}
\hline
Parameter & Prior & Units \\
\hline
\multicolumn{3}{c}{\textbf{Nuclear Matter Parameters (NMPs)}} \\
\hline
$\rho_0$        & U(0.14, 0.17)     & fm$^{-3}$ \\
$E_0$           & U(-16.5, -15.5)   & MeV \\
$K_0$           & U(150, 300)       & MeV \\
$Q_0$           & U(-1500, 400)     & MeV \\
$m^*/m$         & U(0.5, 0.8)       & -- \\
$J_0$           & U(20, 40)         & MeV \\
$L_0$           & U(20, 100)        & MeV \\
\hline
\multicolumn{3}{c}{\textbf{High-Density Parameters (HDPs)}} \\
\hline
$n_{bl}$        & U(0.01, 3.0)      & fm$^{-3}$ \\
$h_p$           & U(0.0, 0.9)       & -- \\
$w_p$           & U(0.1, 5.0)       & fm$^{-3}$ \\
$n_p$           & U(0.08, 5.0)      & fm$^{-3}$ \\
$s_p$           & U(-50, 50)        & -- \\
\hline
\multicolumn{3}{c}{\textbf{Transition Density}} \\
\hline
$\rho_{tr}$        & U(1, 10)          & $\rho_{\text{ref}}$ \\
\hline
\end{tabular}
\end{table}

\noindent
The combined astrophysical likelihood is written as
\begin{equation}
\mathcal{L}_{\rm Astro}
=
\mathcal{L}_{M_{\rm max}}
\times
\mathcal{L}_{\rm NICER}
\times
\mathcal{L}_{\rm GW170817}.
\end{equation}

\noindent
The posterior distributions of neutron star mass and radius inferred from
NICER observations are incorporated into the likelihood using kernel density
estimation (KDE). In particular, we use the reported mass--radius posteriors
for PSR~J0740+6620 ($\sim 2.08\,M_\odot$)~\citep{miller2021radius,riley2021nicer},
PSR~J0030+0451 ($\sim 1.44\,M_\odot$)~\citep{miller2019psr,riley2019nicer},
PSR~J0437--4715 ($\sim 1.418\,M_\odot$)~\citep{Choudhury_2024},
and PSR~J0614--3329 ($\sim 1.44\,M_\odot$)~\citep{Mauviard_2025}.
These KDEs provide the probability density
$P(d_{\rm NICER} \mid m, R)$ entering the likelihood.

\noindent
The NICER likelihood for a given EoS is then defined as
\begin{equation}
 \mathcal{L}_{\rm NICER} =
\int_{M_l}^{M_u} \! {\rm d}m \;
P(m \mid {\rm EOS}) \,
P\!\left(
d_{\rm NICER} \mid
m, R(m, {\rm EOS})
\right)
\label{eq:NICER_likelihood}
\end{equation}
where $P(m \mid {\rm EoS})$ denotes the mass prior implied by the EoS,
$M_l = 1\,M_\odot$ is the lower mass cutoff,
and $M_u = \max(M({\rm EoS}))$ is the maximum mass supported by the EoS.

\noindent
Gravitational-wave observations from the binary neutron star merger GW170817
provide complementary constraints through posterior distributions of the
chirp mass $M_c$ and the combined tidal deformability $\tilde{\Lambda}$.
We fix the chirp mass to the measured value $M_c = 1.186\,M_\odot$ and define
the GW170817 likelihood $\mathcal{L}_{\rm GW}$,  as
\begin{eqnarray}
\mathcal{L}_{\rm GW} &=&
\int_{M_l}^{M_u} \! {\rm d}m_1 \;
P(m_1 \mid {\rm EOS}) \, P\!\left(
d_{\rm GW} \mid
m_1, M_c = 1.186\,M_\odot,
\tilde{\Lambda}(m_1, M_c, {\rm EOS})
\right)
\label{eq:GW_likelihood}
\end{eqnarray}
where $m_1$ is the primary component mass, restricted to the interval
$[1.36, 1.6]\,M_\odot$.

\noindent
Lower bounds on the maximum neutron star mass are imposed using precise radio
timing measurements of massive pulsars.
For each observation $i$, the corresponding likelihood is implemented as a
smoothed step function,
\begin{equation}
\mathcal{L}_{i}(M_{\rm max}) =
\frac{1}{2}
\left[
1 +
\operatorname{erf}
\left(
\frac{M_{\rm max}(\bm{\Theta}) - M_{{\rm max},i}}
{\sqrt{2}\,\sigma_i}
\right)
\right],
\end{equation}
where $M_{{\rm max},i}$ and $\sigma_i$ denote the measured mass and associated
uncertainty of the pulsar.
We include constraints from PSRs J0740+6620~\citep{fonseca2021refined},
J0348+0432~\citep{doi:10.1126/science.1233232},
and J1614--2230~\citep{Arzoumanian_2018}.
The combined maximum-mass likelihood is then given by
\begin{equation}
\mathcal{L}_{M_{\rm max}}(D \mid \bm{\Theta}) =
\prod_i \mathcal{L}_{i}(M_{\rm max}).
\end{equation}

\noindent
In addition to the astrophysical constraints described above, we impose three distinct sets of constraints on the model parameters, corresponding to different approaches for imposing low-density nuclear physics constraints.

\noindent
\textbf{Theoretical constraints:}
This set is defined by two purely theoretical constraints on the EoS, both derived from \emph{ab initio} calculations.
At low densities, chiral effective field theory (EFT) calculations provide constraints on the pressure of $\beta$-equilibrated matter at baryon number densities of
$0.1$, $0.15$, and $0.2~\mathrm{fm}^{-3}$~\citep{PhysRevLett.130.072701}.
At higher densities, additional restrictions are imposed using perturbative QCD (pQCD) calculations~\citep{komoltsev2022perturbative,PhysRevD.109.094030}.
These constraints are implemented by requiring that the energy density $\epsilon$ and pressure ($P$) at $7\rho_{\text{ref}}$ can be causally connected to the asymptotic limits predicted by pQCD.

\noindent
\textbf{Implicit constraints:}
Following Ref.~\citep{tsang2023determination}, we impose constraints on the finite nuclei properties implicitly through ranges of NMPS along with bounds on symmetry energy $J(\rho)$, the symmetry pressure $P_{\rm sym}(\rho)$, and the pressure of symmetric nuclear matter $P_{\rm SNM}(\rho)$. These constraints are derived from analyses of nuclear masses~\citep{Brown2014,Kortelainen2012,Danielewicz2002}, the electric dipole polarizability of ${}^{208}$Pb, and heavy-ion collision (HIC) data across a range of densities. The details of these constraints are provided in Tables~S3 and~S4 of the supplementary material of Ref.~\citep{venneti2024unraveling}.

\noindent
\textbf{Explicit constraints:}
We demonstrated, in Ref.~\citep{venneti2024unraveling}, that constraints placed explicitly on FN constraints influence the properties of NSs. In this, the implicit constraints are contrasted with an explicit treatment of finite nuclei observables.
In this case, the constraints on symmetry energy $J(\rho)$, from analyses of nuclear masses, are replaced by direct constraints on the binding energies and charge radii of
${}^{16}_{8}\mathrm{O}$,
${}^{40}_{20}\mathrm{Ca}$,
${}^{48}_{20}\mathrm{Ca}$, and
${}^{208}_{82}\mathrm{Pb}$.
These observables are computed within the RMF formalism following Refs.~\citep{Ring1997,Gambhir1989}.

The high-density behavior of the EoS beyond a transition density $\rho_{tr}$, a free parameter, is modeled using the speed-of-sound parameterization (Section  \ref{HD}), which serves as a minimal and relatively model-agnostic metamodel without specifying the microscopic composition of matter. In order to ensure physical consistency, we enforce causality,
$(c_s/c)^2 = \mathrm{d}p/\mathrm{d}\epsilon \leq 1$, and require a monotonically increasing pressure,
$\mathrm{d}p/\mathrm{d}\epsilon > 0$, thereby avoiding unphysical NMPs and HDPs.

\section{Results and Discussions}\label{Results_And_Discussions}
We carried out a Bayesian analysis using three distinct sets of constraints on the EoS: theoretical, implicit, and explicit as discussed in the Section~\ref{Bayesian}.
\begin{deluxetable*}{l c c c c c}
\tabletypesize{\scriptsize}
\tablewidth{0pt}
\tablecaption{Median values and 68\% confidence intervals for binding energies ($B$) and charge radii ($R_{ch}$) of selected nuclei obtained using different sets of finite-nuclei constraints (FNCs). Experimental values \citep{Wang_2021,angeli2013table} are shown for reference.\label{tab:FN_prop_table_transposed}}

\tablehead{
\colhead{FNC Set} &
\colhead{Quantity} &
\colhead{${}^{16}_{8}\mathrm{O}$} &
\colhead{${}^{40}_{20}\mathrm{Ca}$} &
\colhead{${}^{48}_{20}\mathrm{Ca}$} &
\colhead{${}^{208}_{82}\mathrm{Pb}$}
}

\colnumbers
\startdata
Theoretical & $B$ &
$-134.08^{+11.17}_{-15.71}$ &
$-352.97^{+23.52}_{-95.63}$ &
$-421.97^{+27.57}_{-398.36}$ &
$-1685.01^{+128.12}_{-5378.19}$ \\
 & $R_{ch}$ &
$2.60^{+0.04}_{-0.08}$ &
$3.33^{+0.06}_{-0.31}$ &
$3.36^{+0.06}_{-0.59}$ &
$5.34^{+0.12}_{-1.36}$ \\
Implicit & $B$ &
$-128.82^{+17.15}_{-14.68}$ &
$-339.96^{+32.24}_{-27.72}$ &
$-403.41^{+36.85}_{-32.85}$ &
$-1593.29^{+126.52}_{-1516.49}$ \\
 & $R_{ch}$ &
$2.68^{+0.04}_{-0.04}$ &
$3.31^{+0.06}_{-0.05}$ &
$3.36^{+0.07}_{-0.05}$ &
$5.35^{+0.12}_{-0.72}$ \\
Explicit & $B$ &
$-128.43^{+0.44}_{-0.44}$ &
$-340.24^{+1.12}_{-1.14}$ &
$-414.49^{+1.23}_{-1.14}$ &
$-1638.17^{+7.20}_{-7.38}$ \\
 & $R_{ch}$ &
$2.703^{+0.006}_{-0.007}$ &
$3.444^{+0.007}_{-0.008}$ &
$3.459^{+0.009}_{-0.008}$ &
$5.517^{+0.016}_{-0.014}$ \\
Experimental & $B$ &
$-127.62$ &
$-342.05$ &
$-416.00$ &
$-1636.43$ \\
 & $R_{ch}$ &
$2.70$ &
$3.48$ &
$3.48$ &
$5.50$ \\
\enddata
\tablecomments{
Binding energies $B$ are in MeV and charge radii $R_{ch}$ are in fm.
}
\end{deluxetable*}

Table~\ref{tab:FN_prop_table_transposed} summarizes the median values and 68\% confidence intervals for the FN observables employed in this work under the three different sets of constraints. A clear trend emerges across the constraint sets for nuclei considered. The explicit constraints yield median binding energies and charge radii that lie closer to the experimental values, accompanied by a significant reduction in the associated uncertainties. This also ensures a uniform description across light, medium, and heavy nuclei. In contrast, both the theoretical and implicit constraint sets result in broad posterior distributions, uncertainties increasing substantially for nuclei with higher mass, particularly for the binding energies. In some cases, the implicit and theoretical constraints permit spreads comparable to the central values themselves, for example $-1593.29^{+126.52}_{-1516.49}$ MeV of  ${}^{208}_{82}\mathrm{Pb}$ for the case of implicit constraints, indicating a loss of predictive power for FN properties.

\begin{table}[htbp]
\centering
\caption{Posterior results for nuclear matter parameters (NMPs), 
high-density parameters (HDPs), and the transition density. 
The width ratio (WR) is defined as 
$\text{width}_{99\%}^{\rm posterior}/\text{width}^{\rm prior}$, where $\text{width}_{99\%}^{\rm posterior}$ is the width of 99\%  confidence interval (CI) and $\text{width}^{\rm prior}$ is the width of the prior.}
\bgroup
\def\arraystretch{1.25}
\label{priors_posteriors}

\begin{tabular}{lccccc}
\hline
\hline
\textbf{Isoscalar NMPs} & $\rho_0$ &
$E_0$ &
$K_0$ &
$Q_0$ &
$m^*/m$ \\
&
(fm$^{-3}$) &
(MeV) &
(MeV) &
(MeV) &
-- \\
\hline
Theoretical &
$0.1595^{+0.0104}_{-0.0187}$ &
$-16.00^{+0.49}_{-0.49}$ &
$235.4^{+78.7}_{-58.5}$ &
$-548^{+758}_{-899}$ &
$0.698^{+0.169}_{-0.101}$ \\
WR &
0.97 & 0.98 & 0.91 & 0.87 & 0.90 \\
\hline
Implicit &
$0.1599^{+0.0099}_{-0.0192}$ &
$-15.99^{+0.48}_{-0.51}$ &
$231.4^{+58.9}_{-73.2}$ &
$-998^{+712}_{-493}$ &
$0.708^{+0.091}_{-0.192}$ \\
WR &
0.97 & 0.99 & 0.88 & 0.63 & 0.95 \\
\hline
Explicit &
$0.1519^{+0.0034}_{-0.0077}$ &
$-15.86^{+0.26}_{-0.41}$ &
$220.2^{+39.5}_{-26.4}$ &
$8.8^{+348}_{-788}$ &
$0.616^{+0.048}_{-0.024}$ \\
WR &
0.37 & 0.67 & 0.44 & 0.60 & 0.24 \\
\hline
\hline
\textbf{Isovector NMPs} &
$J_0$ &
$L_0$ &&&\\
 &
(MeV) &
(MeV) &&&\\
\hline
Theoretical &
$31.85^{+8.01}_{-11.03}$ &
$49.43^{+24.47}_{-22.09}$ \\
WR &
0.95 & 0.58 \\
\hline
Implicit &
$33.22^{+4.41}_{-4.17}$ &
$67.25^{+28.21}_{-27.88}$ \\
WR &
0.43 & 0.70 \\
\hline
Explicit &
$30.36^{+3.76}_{-5.23}$ &
$56.89^{+20.00}_{-16.18}$ \\
WR &
0.45 & 0.45 \\
\hline
\hline
\textbf{HDPs}&
$n_{bl}$ &
$h_p$ &
$w_p$ &
$n_p$ &
$s_p$ \\
 &
(fm$^{-3}$) &
-- &
(fm$^{-3}$) &
(fm$^{-3}$) &
-- \\
\hline
Theoretical &
$0.54^{+2.43}_{-0.52}$ &
$0.29^{+0.58}_{-0.28}$ &
$2.87^{+2.10}_{-2.71}$ &
$2.44^{+2.54}_{-2.34}$ &
$-8.06^{+57.0}_{-41.5}$ \\
WR &
0.99 & 0.95 & 0.98 & 0.99 & 0.99 \\
\hline
Implicit &
$0.28^{+2.67}_{-0.26}$ &
$0.31^{+0.47}_{-0.27}$ &
$2.89^{+2.09}_{-2.65}$ &
$0.69^{+4.21}_{-0.61}$ &
$-4.14^{+53.6}_{-45.3}$ \\
WR &
0.98 & 0.83 & 0.97 & 0.98 & 0.99 \\
\hline
Explicit &
$0.37^{+2.58}_{-0.36}$ &
$0.31^{+0.35}_{-0.21}$ &
$2.64^{+2.39}_{-2.37}$ &
$0.62^{+4.13}_{-0.52}$ &
$-5.99^{+55.5}_{-43.4}$ \\
WR &
0.98 & 0.62 & 0.96 & 0.95 & 0.99 \\
\hline
\hline
\textbf{Transition Density}&
$\rho_{tr}$ \\
& ($\rho_{\text{ref}}$) \\
\hline
Theoretical &
$2.66^{+7.19}_{-1.62}$ \\
WR & 0.98 \\
\hline
Implicit &
$2.02^{+7.21}_{-1.01}$ \\
WR  & 0.91 \\
\hline
Explicit &
$1.92^{+2.67}_{-0.92}$ \\
WR & 0.40 \\
\hline
\hline
\end{tabular}
\egroup
\end{table}

\begin{figure}[h!]
    \centering
    \includegraphics[width=\textwidth]{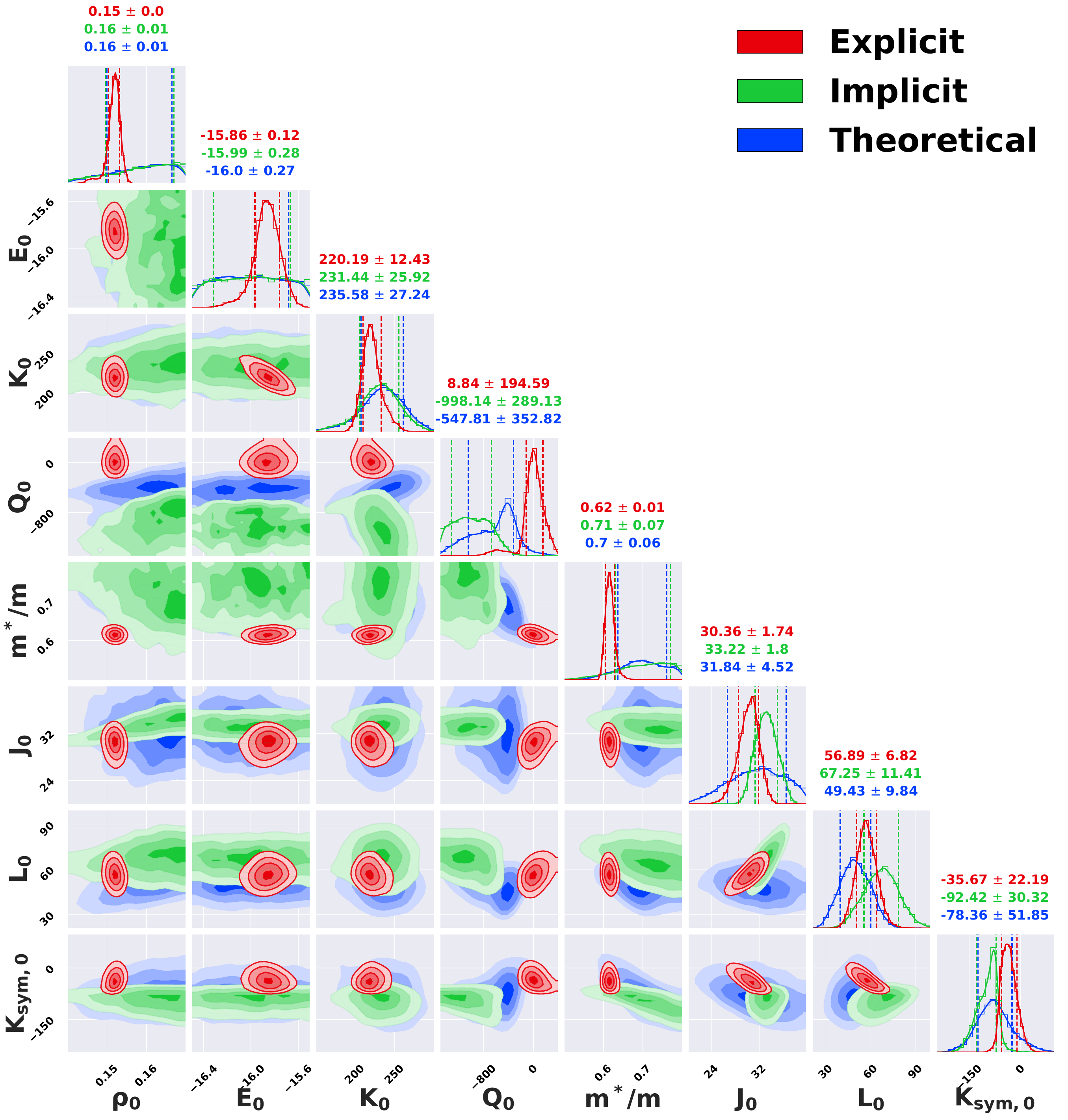}
    \caption{Posterior distributions of nuclear matter parameters of the RMF model for the three distinct set of constraints: \textit{Theoretical}, \textit{Implicit} and \textit{Explicit}. The 1$\sigma$ intervals are shown as vertical dashed lines in the marginalized posterior distributions. All are in the units of MeV, except $\rho_0$ and $m^*/m$ which are in fm$^{-3}$ and dimensionless respectively.}
    \label{fig:NMPs_corner_Tr2}
\end{figure}

Fig.\ref{fig:NMPs_corner_Tr2} presents the posterior distribution of the NMPs for the three sets of constraints. 
The \textit{theoretical} constraints yield the posterior distributions with the largest uncertainties, as evidenced by broader widths across most of the parameters. Particularly, $\rho_0$ and $E_0$ remain largely unconstrained, with no noticeable reduction in their widths relative to the priors. The \textit{theoretical} case also favors relatively higher values for $K_0$ and lower values of $L_0$ compared to the other sets. The set of \textit{implicit} constraints display a similarly limited narrowing of $\rho_0$ and $E_0$. Overall, this case tends to favor larger values for $m^*/m$, $J_0$ and $L_0$, along with smaller values of $K_{sym,0}$ and $Q_0$. The distribution of $K_0$ remains comparable to that obtained from the theoretical constraints. These sets are contrasted with the case of explicit constraints. Explicit constraints lead to a substantial reduction in uncertainties across the parameter space. The most striking improvement is for $m^*/m$, whose posterior width is reduced by a factor of six relative to the previous cases. While the uncertainty in $J_0$ marginally reduces as compared to implicit case, the explicit set prefers lower values of both $J_0$ and $m^*/m$, together with higher values for $K_{sym,0}$ and $Q_0$. The posterior for $L_0$ lies between those of implicit and theoretical cases, a region reasonably consistent with both. This significant reduction in the uncertainties reflects the strong influence of FN observables, whose precise measurements permit only a narrow range of plausible combinations of the NMPs. The three sets of distributions also exhibit distinct inter-parameter correlations. This is most evident for the pair ($E_0$, $K_0$).  There is no discernible correlation between them for theoretical and implicit case, but displays pronounced anti correlation for explicit FN constraints. As emphasized in earlier studies\citep{Malik2020, Patra2023, venneti2024unraveling}, such correlation structure are essential in interpreting the the relation of the NMPs with NS observables. The width ratio, defined as $\text{width}_{99\%}^{\rm posterior}/\text{width}^{\rm prior}$, where $\text{width}_{99\%}^{\rm posterior}$ is the width of 99\%  confidence interval (CI) and $\text{width}^{\rm prior}$ is the width of the prior quantifies the reduction in the posterior widths in comparison with priors. This ratio along with the posterior medians and 99\% CI are shown in Table.\ref{priors_posteriors}.

\begin{figure}[h!]
    \centering
    \includegraphics[width=0.6\textwidth]{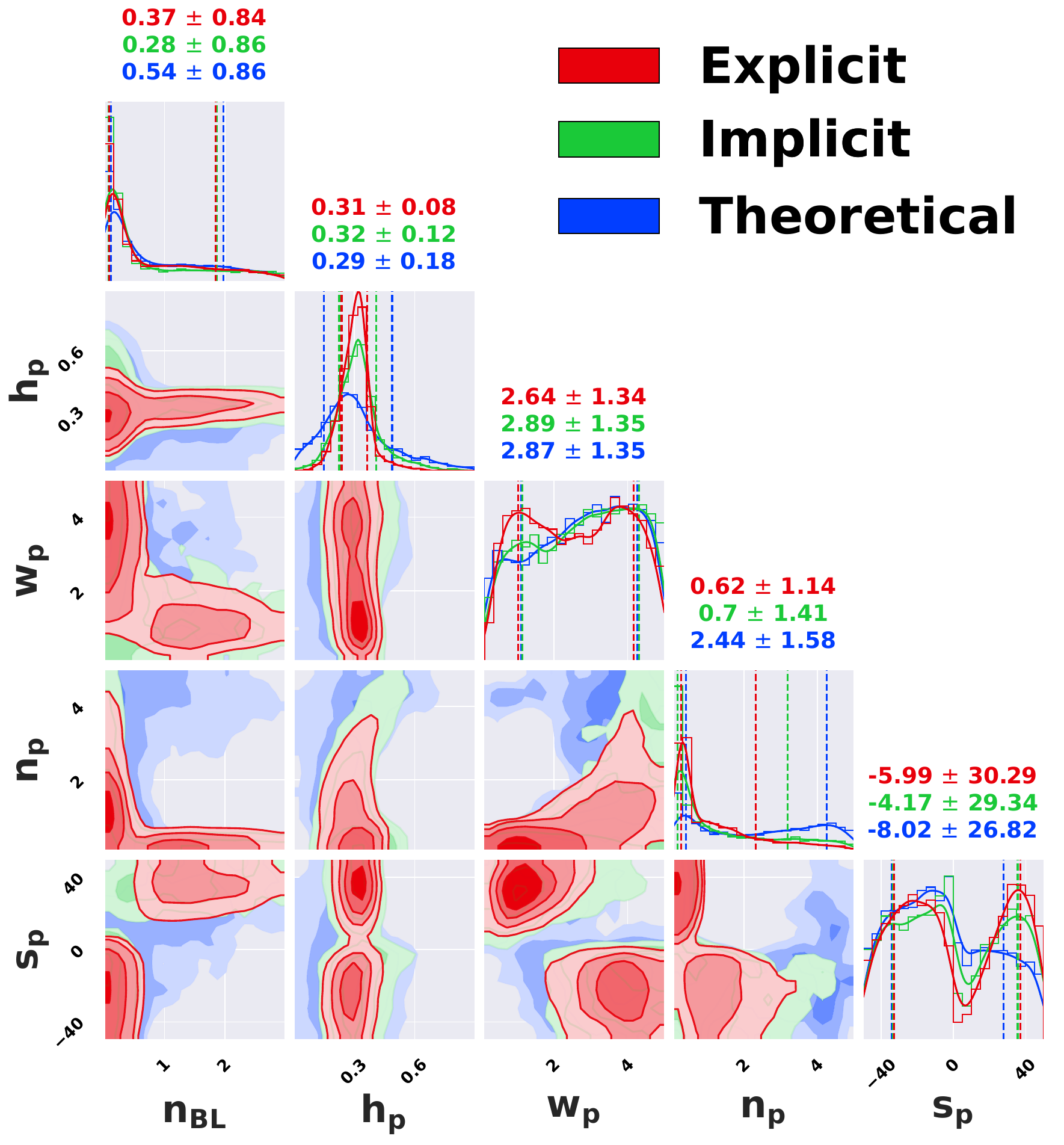}
    \caption{Posterior distributions for high density parameters. All are in the units of fm${}^{-3}$,  except $h_p$ and $s_p$ which are dimensionless.}
    \label{fig:HDPs_corner_Tr2}
\end{figure}

In our analysis, the high density EoS, above the transition density $\rho_{tr}$, is generated using the speed of sound parameterization of \citet{Tews2018}. Fig.\ref{fig:HDPs_corner_Tr2} shows the posterior distributions of HDPs. The median values, along with 99\% confidence intervals, are also listed in Table \ref{priors_posteriors}. The HDPs display broadly similar trends across all the three sets of constraints. The only parameter that is meaningfully constrained is the peak height, $h_p$, whose posterior width decreases from the theoretical case to explicit case. This is also evident in the reduction of  width ratio from $0.95$ to $0.63$, as presented in Table.\ref{priors_posteriors}. The rest of the parameters remain largely unconstrained. For instance, $w_p$ and $s_p$ retain broad distributions while $n_{BL}$ and $n_p$ exhibit long tailed posteriors. The implicit and explicit constraints tend to favor lower values of $n_p$ and $n_{BL}$, which are significantly smaller than the ones for theoretical constraints. The skewness parameter $s_p$, often fixed to zero in earlier works \citep{Patra2023}, develops multiple peaks on the either side of zero as observed in the marginal posterior distribution in Fig.\ref{fig:HDPs_corner_Tr2}.

\begin{figure}[h!]
    \centering
    \includegraphics[width=0.6\textwidth]{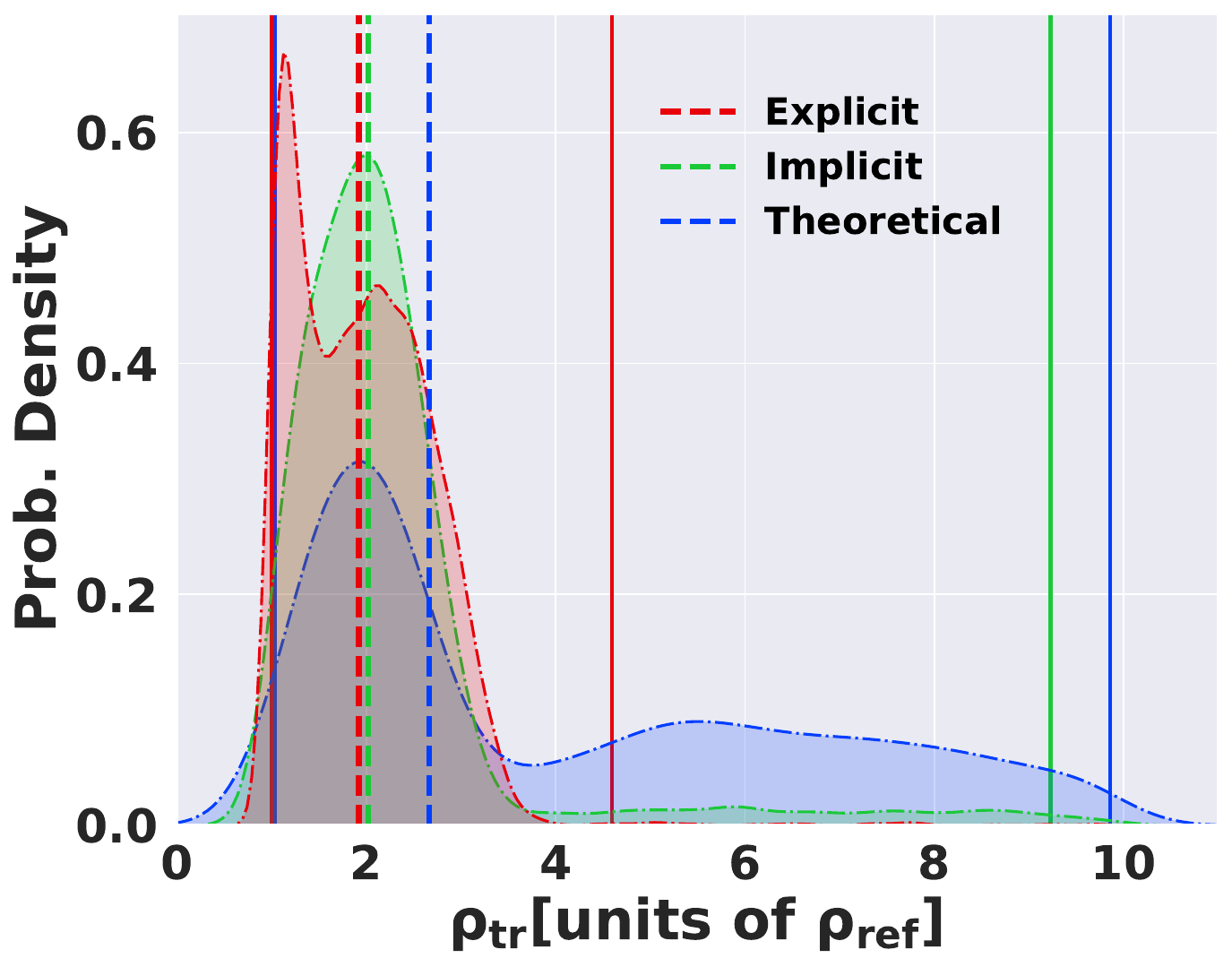}
    \caption{Posterior distributions of the transition density $\rho_{tr}$ for the three sets of distinct constraints. Vertical dashed lines are the median values while the solid lines bound the 99\% CI}
    \label{fig:NTR_Distribution}
\end{figure}

The probability density of $\rho_{\mathrm{tr}}$ for the three cases are plotted in Fig.~\ref{fig:NTR_Distribution}. Interestingly, all three constraint sets yield posterior distributions that peak near $2\rho_{\text{ref}}$, albeit with markedly different tail behaviors. The posterior width of $\rho_{\mathrm{tr}}$ decreases progressively from the theoretical to the implicit and finally to the explicit constraints, with corresponding width ratio $\sim 0.98$,$\sim 0.91$ and $\sim 0.4$ respectively (Table.\ref{priors_posteriors}), reflecting a substantial reduction in the uncertainties in $\rho_{tr}$ in the explicit case. While the theoretical and implicit sets allow transition densities over a broad range of $\sim 1-10\rho_{\text{ref}}$, the explicit constraints restrict this to only $\sim 1-4\,\rho_{\text{ref}}$. This pronounced restriction in the allowed transition region underscores the strong influence of low-density FN constraints in determining the onset of possible phase transitions in neutron-star matter. 

\begin{figure}[h!]
    \centering
    \includegraphics[width=\textwidth]{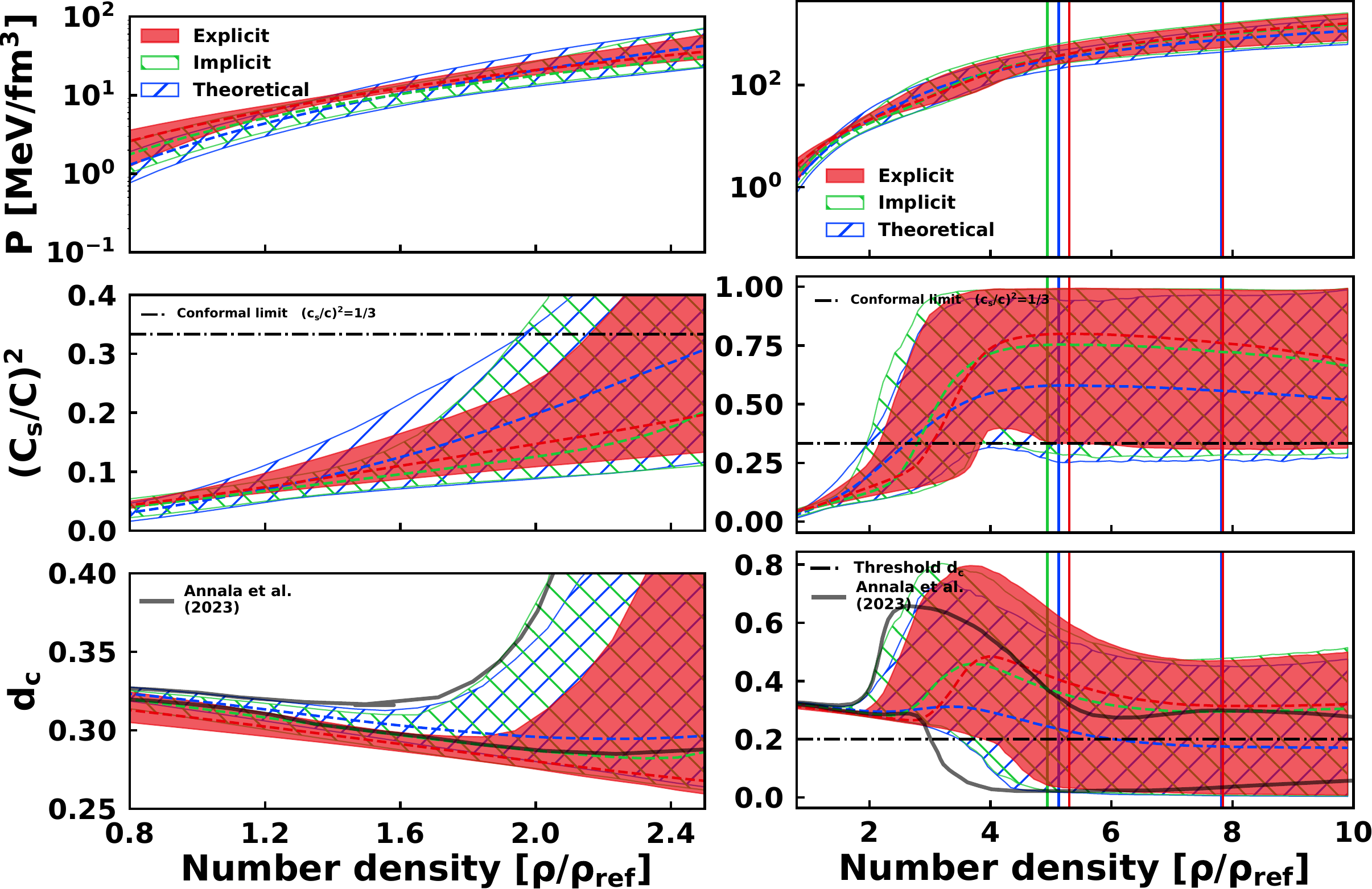}
    \caption{99\% confidence regions for pressure $P$, squared speed of sound $(c_s/c)^2$, and the conformality parameter $d_c$ across the low densities (0.8-2.5 $\rho_{\text{ref}}$ in the left panels), and across the density range upto 10$\rho_{\text{ref}}$ in the right panels. Also shown are the 95\% credible region from \citep{annala2023strongly} bounded by thick black lines in (e) and (f), along with the speed of sound in the conformal limit in (c) and (d) shown as horizontal dash-dotted lines. The vertical lines cover the 99\% CI of the central density, corresponding to maximum mass stars for each of the case. The threshold conformality parameter $d_c = 0.2$, which indicates the proximity to conformal limit  is also shown as dash-dotted lines in (f).}
    \label{fig:EoS_distr}
\end{figure}

Next, we present the $99\%$ credible intervals (CIs) of the pressure, the squared speed of sound $c_s^2/c^2$, and the conformality parameter $d_c$ across the three constraint sets and over different density regimes in Fig.~\ref{fig:EoS_distr}(a--f). In the left panel, the low density region is zoomed while the right panel shows the plots across the entire range of densities. The pressure distributions are shown in top row of Fig.~\ref{fig:EoS_distr}. In the low density ($0.8 - 1.0 \, \rho_{\text{ref}}$) region, the explicit constraint set yields the stiffest EoS up to densities of $\sim 1.5\,\rho_0$, while the theoretical set remains comparatively softer, as reflected in the median pressure curves (shown as dashed lines). Beyond $1.5\,\rho_{\text{ref}}$, the stiffness of the explicit and theoretical cases becomes comparable, with the implicit case remaining slightly softer. Throughout the low-density regime, the pressure uncertainty is narrowest for the explicit finite-nucleus (FN) constraints and widest for the theoretical constraints. At higher densities beyond $2.5\,\rho_{\text{ref}}$ (Fig.~\ref{fig:EoS_distr}b), the theoretical case remains marginally stiffer till $4\,\rho_{\text{ref}}$ as compared to the implicit and explicit cases, which show similar stiffness. In this regime, the pressure uncertainties become comparable across all three cases, with a noticeable broadening for the explicit constraints around $2.5\,\rho_{\text{ref}}$.

The behavior of the squared speed of sound is shown in the middle panels of Fig.~\ref{fig:EoS_distr}. At densities below $\sim1.2\,\rho_{\text{ref}}$, the theoretical case exhibits the lowest median values of $c_s^2/c^2$, which subsequently rises and overtakes the other cases at higher densities. In the intermediate density range ($1$--$2.5\,\rho_{\text{ref}}$), the implicit case is comparatively softer. However beyond $2.5\,\rho_{\text{ref}}$ it stiffens rapidly and marginally exceeds the other two cases up to $\sim 4\,\rho_{\text{ref}}$. The explicit case shows the highest values at densities below $1\,\rho_{\text{ref}}$, followed by a steady increase that crosses the implicit case near $5\,\rho_{\text{ref}}$. Both the implicit and explicit cases exhibit a rapid rise in $c_s^2/c^2$ over a narrow density interval of $1$--$1.5\,\rho_{\text{ref}}$, after which they gradually approach the conformal limit ($c_s^2/c^2 = 1/3$). In contrast, the theoretical case displays a more gradual increase toward a lower peak before exhibiting a similar decline.

In the bottom panel of Fig.\ref{fig:EoS_distr}, we plot the conformality parameter defined as\citet{annala2023strongly, fujimoto2022trace}
\[
d_c = \sqrt{\Delta^2 + (\Delta')^2},
\]
where $\Delta$ is the renormalized trace anomaly \citep{}. Values of $d_c \leq 0.2$ indicate proximity to the conformal limit and may signal the presence of quark matter. The corresponding reference regions from \citet{annala2023strongly} (bounded by thick black lines) are overlaid on the $99\%$ CIs of the three constraint sets in Fig.~\ref{fig:EoS_distr}(e,f). At low densities (Fig.~\ref{fig:EoS_distr}e), all cases yield $d_c \gtrsim 0.2$, with median values near $0.3$--$0.35$, and the implicit and explicit cases lying largely outside the reference region. At higher densities, the uncertainties increase and the overlap between cases becomes more pronounced, although the distributions remain significantly skewed toward lower values.

Following the trend of $d_c$ over the entire range of densities (Fig.~\ref{fig:EoS_distr}f), the theoretical constraint set shows better agreement with the reference region at lower densities, while the implicit and explicit cases exhibit a rapid rise in $d_c$ around $2.5\,\rho_{\text{ref}}$ over a density interval of $1$--$1.5\,\rho_{\text{ref}}$, followed by a gradual decline beyond a peak near $4\,\rho_{\text{ref}}$. At densities above $6\,\rho_{\text{ref}}$, the implicit and explicit cases plateau, with substantial portions of their posteriors still lying outside the reference region. Although all three cases display a peak qualitatively similar to that reported by \citet{annala2023strongly}, the peak appears shifted to higher densities, around $4\,\rho_{\text{ref}}$, and is broader for the theoretical constraints. Among the three cases, only the theoretical set exhibits a significant posterior region with $d_c \leq 0.2$.

\begin{figure}[h!]
    \centering
    \includegraphics[width=0.75\textwidth]{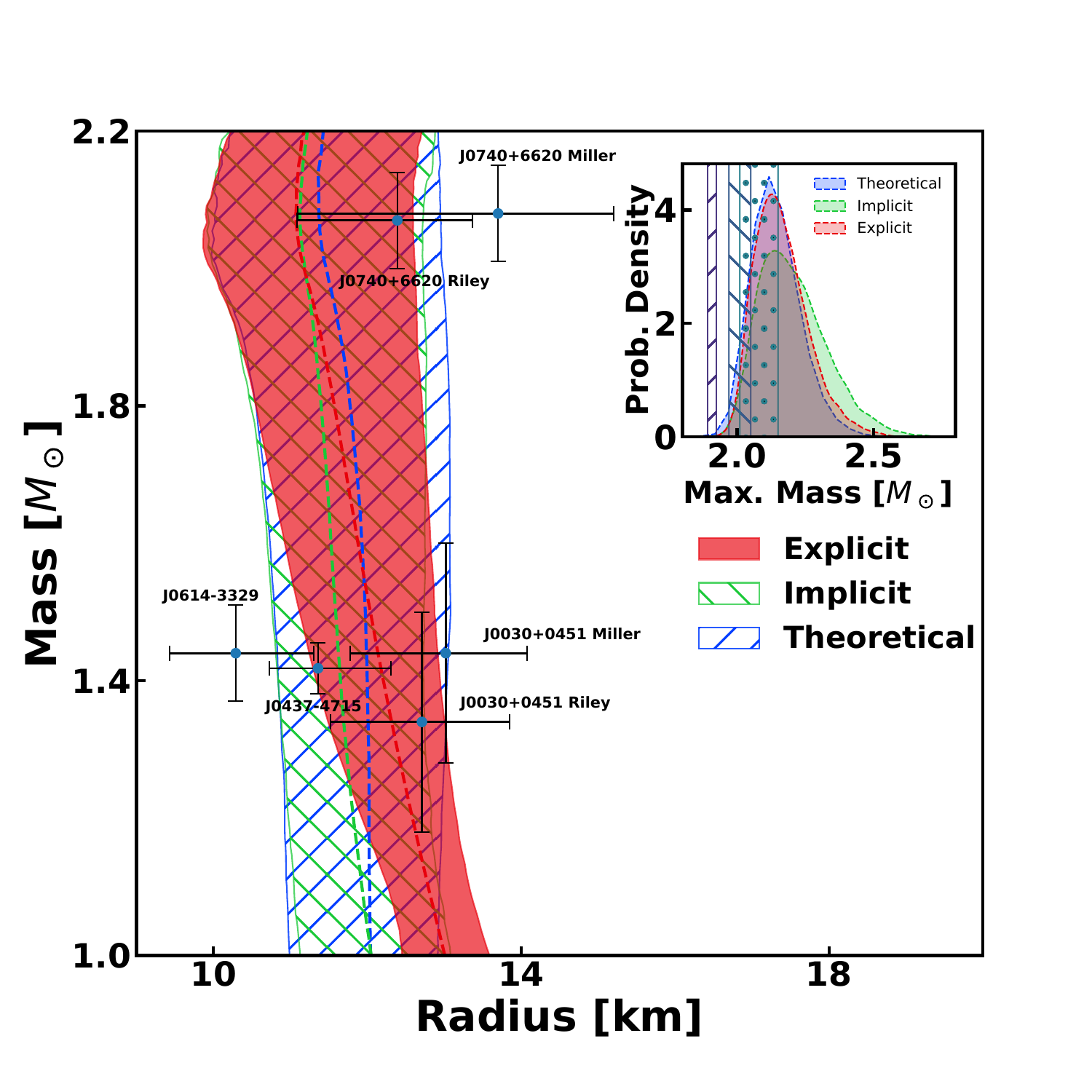}
    \caption{99\% confidence regions for mass and radius for the three distinct cases. The observed mass-radius of the recent pulsar observations are shown are shown. (inset) The maximum mass distribution for the three cases along with the observational lower limit on the maximum mass coming from PSRs J0740+6620~\citep{fonseca2021refined} (dotted patch),
J0348+0432~\citep{doi:10.1126/science.1233232} (left hatched),
and J1614--2230~\citep{Arzoumanian_2018} (right hatched).}
    \label{fig:MR_Distribution}
\end{figure}

Having discussed the properties of the EoS distributions, we focus on the trends emerging in the NS properties. The $99\%$ credible regions of the NS mass--radius ($M$--$R$) distributions for the three constraint sets are shown in Fig.~\ref{fig:MR_Distribution}. The median mass--radius estimates inferred from NICER observations of PSRs J0437--4715, J0614--3329, J0030+0451, and J0740+6620 are overlaid for reference. At the high-mass end, all three cases exhibit broadly similar $M$--$R$ distributions, with only minor differences in their median trends. The distributions of the maximum mass (inset of Fig.~\ref{fig:MR_Distribution}) also show comparable behavior across the three cases, which is consistent with the observed pattern in the posterior distributions of the HDPs.

In contrast, differences between the constraint sets become more pronounced in the low-mass region of the $M$--$R$ plane, reflecting the impact of low-density EoS constraints. In particular, the explicit constraint set yields systematically larger median radii at low masses, pushing its $M$--$R$ region further from the NICER measurements of PSRs J0614--3329 and J0437--4715. This offset suggests a potential tension between the explicit finite-nucleus constraints and the mass--radius observations of these sources.

To quantify the agreement between the EoS-generated mass--radius curves and the NICER observations, we compute the Mahalanobis distance with respect to each observational posterior characterized by a mean vector $\boldsymbol{\mu}$ and covariance matrix $\boldsymbol{\Sigma}$ \citep{mahalanobis2018generalized, johnson2002applied},
\begin{equation}
D_M(\boldsymbol{x}) = \sqrt{(\boldsymbol{x} - \boldsymbol{\mu})^{\mathrm{T}} \boldsymbol{\Sigma}^{-1} (\boldsymbol{x} - \boldsymbol{\mu})}.
\end{equation}
This metric accounts for correlations between mass and radius and provides a natural measure of statistical separation in the anisotropic observational parameter space.

We summarize the comparison in Table~\ref{table: MR} using two complementary metrics. The first is the minimum Mahalanobis distance (\textbf{MinMD}), which quantifies how closely a given constraint set allows any portion of the mass--radius curve to approach the observational posterior. The second metric, \textbf{FracInside}, measures the fraction of the full mass--radius curve that lies within a specified observational confidence region (68\% or 95\%). While MinMD captures the ability of a model to reach the allowed region, FracInside reflects its overall consistency with the observational uncertainty contours. The quantitative comparison using the Mahalanobis-distance–based metrics reveals a clear distinction in the constraining power of the NICER observations across the three sets of constraints. For high-mass pulsars, such as PSR J0740+6620, all three constraint sets exhibit comparable agreement, both in terms of their ability to reach the observational posterior (MinMD) and in their geometric overlap with the confidence contours (FracInside). This consistency reflects the similarity of the high-density EoS behavior across the three sets, as already indicated by the posterior distributions of the HDPs, as well as the mass-radius distributions.

In contrast, significant differences emerge for low-mass pulsars, most notably PSR J0614--3329. In this cases, the explicit FN constraints exhibit reduced MinMD values and FracInside fractions at the $68\%$ confidence level, with higher values at $95\%$. This behavior indicates that EoSs satisfying the explicit constraints struggle to approach them closer than $68\%$ confidence level, while having significantly lesser fraction of the curves lying inside, as comapared to the other cases. The theoretical and implicit constraint sets, by comparison, retain substantially better agreement with the low-mass observations. These results suggest a possible systematic tension between the explicit FN constraints and low-mass neutron-star mass--radius measurements at present uncertainties. 

\begin{table}[h!]
\centering
\begin{tabular}{lcccccc}
\hline
\textbf{Source} & \multicolumn{3}{c}{\textbf{MinMD}} & \multicolumn{3}{c}{\textbf{FracInside}} \\
& Theo & Impl & Expl & Theo & Impl & Expl \\
\hline
\multicolumn{7}{l}{\textbf{J0740+6620}} \\
$\;\;$68\% & 0.7374 & 0.6545 & 0.6055 & 0.1885 & 0.1418 & 0.1348 \\
$\;\;$95\% & 0.9913 & 0.9947 & 0.9983 & 0.4851 & 0.4829 & 0.4835 \\
\hline
\multicolumn{7}{l}{\textbf{J0030+0451}} \\
$\;\;$68\% & 0.8849 & 0.9043 & 1.0000 & 0.214 & 0.1819 & 0.2122 \\
$\;\;$95\% & 0.9928 & 0.9965 & 1.0000 & 0.3914 & 0.3495 & 0.3628 \\
\hline
\multicolumn{7}{l}{\textbf{J0437--4715}} \\
$\;\;$68\% & 0.9420 & 0.9888 & 0.9445 & 0.3798 & 0.444 & 0.3497 \\
$\;\;$95\% & 0.9927 & 0.9963 & 0.9999 & 0.7297 & 0.7612 & 0.7095 \\
\hline
\multicolumn{7}{l}{\textbf{J0614--3329}} \\
$\;\;$68\% & 0.4015 & 0.6483 & 0.2629 & 0.0812 & 0.1217 & 0.0379 \\
$\;\;$95\% & 0.9394 & 0.988 & 0.9614 & 0.3294 & 0.3957 & 0.3029 \\
\hline
\end{tabular}
\caption{\label{table: MR}
Quantitative comparison of EoS-generated mass--radius curves with NICER observations.
\textbf{MinMD} denotes the fraction of posterior samples whose minimum Mahalanobis distance lies within the specified confidence region, while \textbf{FracInside} represents the fraction of the mass--radius curve enclosed by the observational contour.
Higher values indicate stronger agreement.
}
\end{table}

\begin{figure}[h!]
    \centering
    \includegraphics[width=0.75\textwidth]{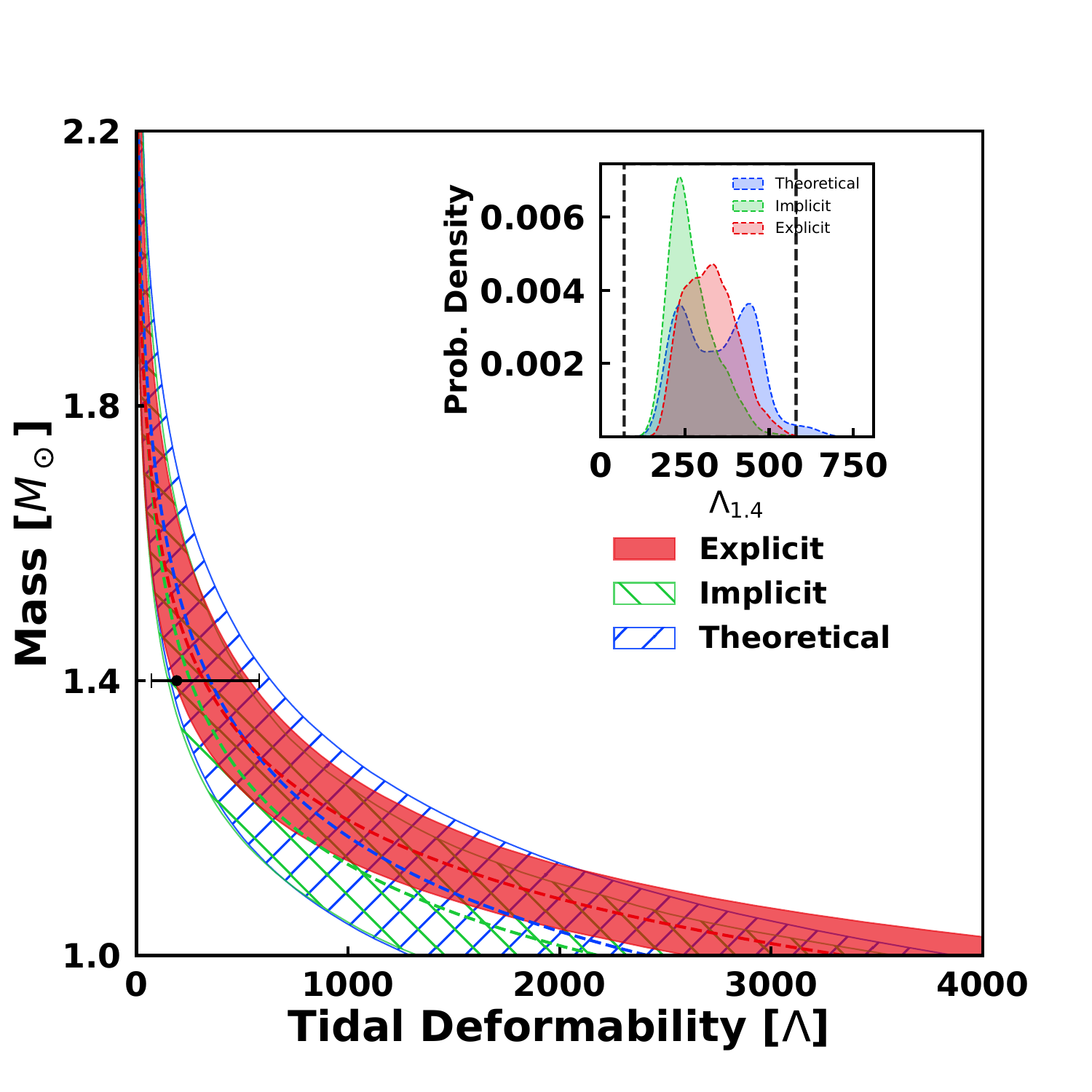}
    \caption{99\% credible region for mass tidal deformability for the three different cases. The observational value (95\% CI) of canonical dimensionless tidal deformability from GW170817 is also shown as a horizontal black bar ($190^{+390}_{-190}$) at 1.4$M_\odot$. (inset) The distribution of canonical tidal deformability is shown along with the observational limits (95\% CI) as bounded by dashed black lines.}
    \label{fig:MT_Distribution}
\end{figure}

In addition to the mass--radius distributions, measurements of tidal deformability from GW170817 provide important and complementary constraints on the neutron-star EoS. Figure~\ref{fig:MT_Distribution} shows the 99\% credible intervals for the mass--dimensionless tidal deformability relation across the three sets of constraints. As expected, the tidal deformability decreases towards high mass NSs. While all three cases yield similar predictions at higher masses, their behavior diverges significantly in the low-mass regime. Among the three cases, the implicit constraints predict the lowest tidal deformabilities at low masses, whereas the explicit constraints favor higher values. In our earlier work \citep{venneti2024unraveling}, we reported a relatively high canonical tidal deformability lying near the upper edge of the observationally inferred interval with minimal overlap. With the introduction of a generalized high-density EoS parametrization, the explicit constraint set now acquires sufficient flexibility to yield canonical tidal deformabilities consistent with the observed range (see inset of Fig.~\ref{fig:MT_Distribution}). An additional noteworthy feature is the bimodal structure observed in the tidal deformability posterior for the theoretical constraint set, which is absent in both the implicit and explicit cases. This behavior reflects the freedom in the high-density EoS permitted under purely theoretical constraints. Although the explicit constraints are now largely consistent with observational bounds, they continue to favor marginally higher canonical tidal deformability values compared to the other two cases.

\begin{figure}[h!]
    \centering
    \includegraphics[width=0.65\textwidth]{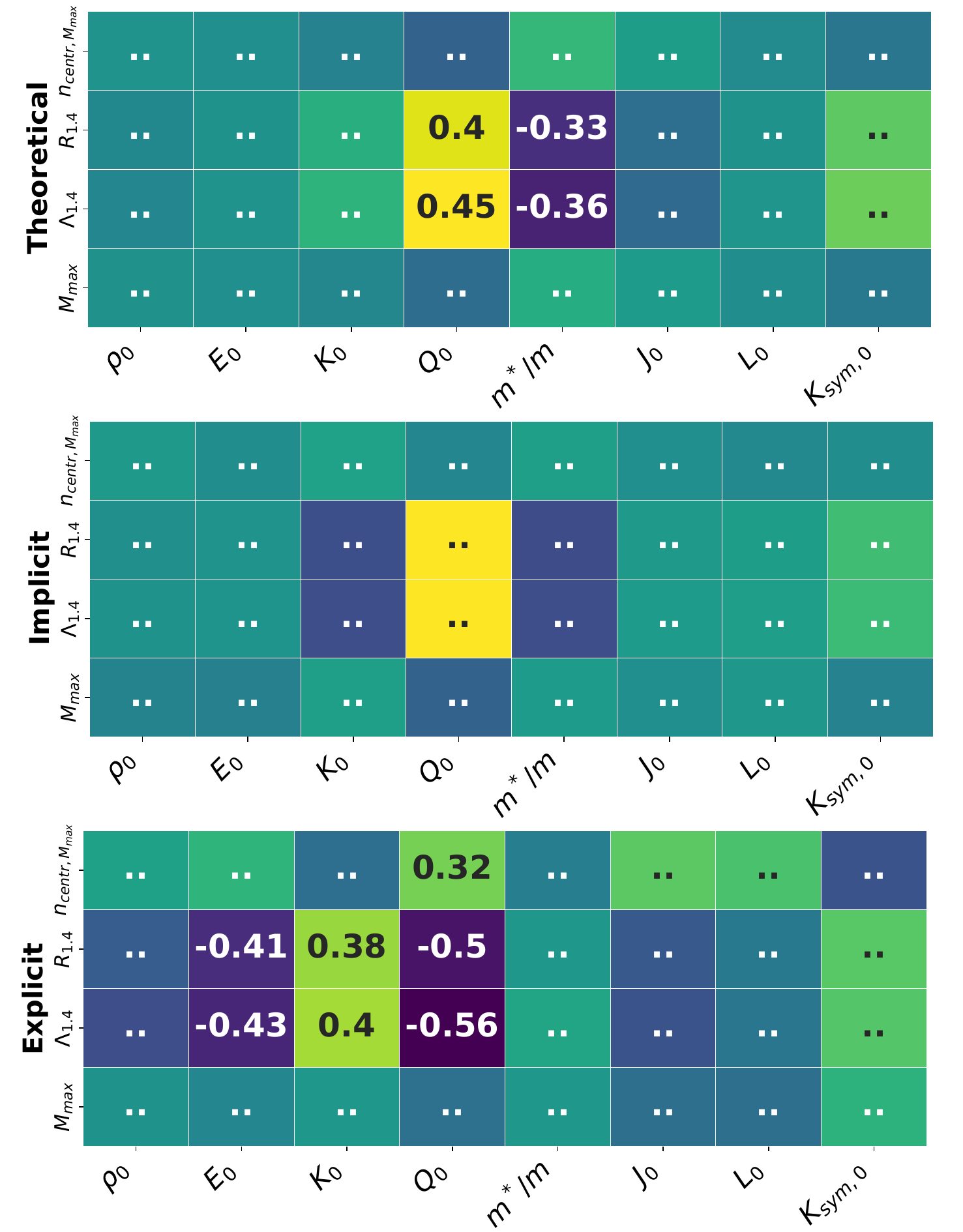}
    \caption{ The correlation heat map between the nuclear matter parameters and the neutron star properties for theoretical, implicit and explicit constraints. The correlations with absolute values >0.3 are displayed. The negative (positive) correlation is shown in the boxes with shades of blue (yellow), with darker shades indicating a higher negative (positive) correlation.}
    \label{fig:Correlation}
\end{figure}

Finally, Fig.~\ref{fig:Correlation} presents the correlation heatmaps for the three sets of constraints, only displaying the Pearson correlation coefficient over 0.3. The correlation structures differ markedly across the cases, primarily due to differences in inter-parameter correlations at low densities. In addition, the inclusion of high-density parameters that govern the EoS beyond the transition density $\rho_{tr}$ further modifies these correlations.

Under the theoretical constraints, the canonical radius and tidal deformability show weak to moderate correlations, particularly with $Q_0$ and $m^*/m$.  In contrast, the implicit case exhibits no statistically significant correlations. The explicit case displays a comparatively richer correlation structure; however, the correlations remain moderate in magnitude. 
Notably, the weak correlation between $Q_0$ and $\Lambda_{1.4}$ seen in the theoretical case transitions to a moderate anti-correlation when explicit constraints are imposed. Another salient feature is the general absence of correlations between the symmetry-energy parameters and the canonical tidal deformability across all three cases. 
In our earlier work~\citep{venneti2024unraveling}, a purely nucleonic RMF description subject to explicit constraints exhibited stronger and more diverse correlation patterns than those obtained under implicit constraints. The differences in the patterns observed here arise mainly from the introduction of a generalized high-density EoS, wherein the RMF description is restricted to densities below $\rho_{tr}$. This behavior underscores the sensitivity of correlation patterns to both the treatment of low-density nuclear physics and the assumed freedom of the high-density EoS.
 
\section{Conclusions}\label{Conclu}

We performed a comprehensive Bayesian analysis of the neutron star (NS) equation of state (EoS) using three distinct sets of constraints on NS EoS, \textit{theoretical}, \textit{implicit}, and \textit{explicit} in addition to recent astrophysical observations. The low density part is constructed using the relativistic mean field model up to a transition density $\rho_{\mathrm{tr}}$. Beyond this, we use a speed-of-sound parameterization, capturing a generalized high-density behavior. This framework enabled us to systematically disentangle the impact of low-density nuclear physics from assumptions about the composition and dynamics of matter at extremely high densities. The inferred transition densities should hence be interpreted as phenomenological indicators within our adapted framework, rather than direct evidence for any specific microscopic phase transition.

Our results demonstrate that explicit finite nuclei (FN) constraints exert the strongest influence on the nuclear matter parameters. They substantially reduce the uncertainties in the posterior distributions, most notably for the effective mass $m^*/m$. In contrast, the theoretical and implicit constraint sets leave several low-density parameters, such as $\rho_0$ and $E_0$, largely unconstrained.

While all three constraint sets favor $\rho_{\mathrm{tr}} \sim 2\rho_{\text{ref}}$, the explicit constraints significantly suppress the high-density tail, restricting the transition to occur in the range $\sim 1-4\,\rho_{\text{ref}}$,as compared to $\sim 1-10\,\rho_{\text{ref}}$ of the other cases. This highlights the critical role of low density explicit FN constraints in limiting the onset of possible non-nucleonic degrees of freedom in neutron-star matter.

The EoSs satisfying explicit FN constraints predict systematically larger radii at low masses, leading to a reduced agreement with NICER observation of PSR J0614--3329. A quantitative comparison using Mahalanobis-distance–based metrics indicates a possible tension (within 68\% confidence regions) between the explicit FN constraints and these low-mass observations, whereas the theoretical and implicit cases display no such tension.

The analysis of tidal deformability from GW170817 provides complementary insights. At higher masses, all three cases yield similar predictions. Differences again emerge in the low-mass regime. While explicit constraints previously in the case of purely nucleonic matter \citep{venneti2024unraveling} favored larger canonical tidal deformabilities, the generalized high-density EoS allows sufficient flexibility for them to remain within the observational bounds. However, the median tidal deformability of the posteriors of explicit constraints has a marginally higher value in comparison to those of implicit constraints.

Finally, we studied the correlations of the NMPs with NS observables. These reveal the sensitivity of inferred correlations to low density constraints as well as the modeling of high density physics. Correlations that are weak under theoretical constraints are significantly strengthened, or even change sign, when FN constraints are imposed explicitly. In contrast, the symmetry energy parameters exhibit no significant correlation with canonical tidal deformability in all the cases. These results underscore that the inferred correlations between NMPs and NS observables are not universal, but heavily model- and constraint-dependent.

Overall, our study emphasizes the dominant role of low-density nuclear physics in shaping NS radii, transition densities, and observable correlations, even with a flexible high-density EoS. The emerging tension between explicit finite-nucleus constraints and low-mass neutron-star observations points to either residual systematic uncertainties in current measurements or missing physics in the modeling of dense matter, motivating further theoretical and observational efforts.
\section*{Acknowledgements}
BKA acknowledges with gratitude the support received under the Raja Ramanna Chair scheme of the Department of Atomic Energy (DAE), Government of India. AV acknowledges fruitful discussions with Debanjan Guha Roy and Skund Tewari. AV also acknowledges financial support from the CSIR--HRDG through the CSIR--JRF grant No.~09/1026(16303)/2023--EMR--I.
\bibliography{References}{}
\bibliographystyle{aasjournalv7}



\end{document}